\documentclass[usegraphicx,usenatbib,fleqn]{mn2e}
\usepackage{mathptmx}

\usepackage{amssymb,amsmath}
\usepackage[dvipdfm]{hyperref}
\usepackage{aas_macros}

\usepackage[letterpaper,margin=0.8in]{geometry}
\newcommand{\im}{\ensuremath{\overline{m}}} 
\newcommand{\valunc}[2]{\ensuremath{#1\pm#2}} 
\newcommand{\valunca}[3]{\ensuremath{#1_{-#2}^{+#3}}} 
\newcommand{\nupeak}{\ensuremath{\nu_{\mathrm{pk}}}}
\newcommand{\nodata}{...}

\begin{document}

\title[Radio variability of gamma-ray blazars]{Connecting radio variability to the characteristics of gamma-ray blazars}

\author[J.~L. Richards et al.]{J.~L.~Richards,$^1$\thanks{E-mail: jlr@purdue.edu} T.~Hovatta,$^2$
  W.~Max-Moerbeck,$^3$ V.~Pavlidou,$^{4,5}$ T.~J.~Pearson,$^2$
  \newauthor
  and A.~C.~S.~Readhead$^2$\\
$^1$Department of Physics, Purdue University, 525 Northwestern Ave, West Lafayette, IN  47907\\
$^2$Cahill Center for Astronomy and Astrophysics, California Institute of Technology,
  1200~E~California Blvd, Pasadena, CA 91125\\
$^3$National Radio Astronomy Observatory, PO Box 0, Socorro, NM 87801\\
$^4$Foundation for Research and Technology -- Hellas, IESL, Voutes, 7110 Heraklion, Crete, Greece\\
$^5$Department of Physics and Institute of Theoretical \& Computational Physics, University of Crete, PO Box 2208, GR-710 03, Heraklion, Crete, Greece}

\maketitle
\begin{abstract}
  We present results from four years of twice-weekly 15~GHz radio monitoring of about 1500 blazars
  with the Owens Valley Radio Observatory 40~m telescope.  Using the intrinsic modulation index to
  measure variability amplitude, we find that, with $>\!6\sigma$ significance, the radio variability
  of radio-selected gamma-ray-loud blazars is stronger than that of gamma-ray-quiet blazars.  Our
  extended data set also includes at least 21~months of data for all AGN with `clean' associations
  in the \emph{Fermi} Large Area Telescope First AGN catalogue, 1LAC. With these additional data we
  examine the radio variability properties of a gamma-ray-selected blazar sample. Within this
  sample, we find no evidence for a connection between radio variability amplitude and optical
  classification. In contrast, for our radio-selected sample we find that the BL~Lac object
  subpopulation is more variable than the flat spectrum radio quasar (FSRQ) subpopulation. Radio
  variability is found to correlate with the synchrotron peak frequency, with low- and
  intermediate-synchrotron-peaked blazars varying less than high-synchrotron-peaked ones. We find
  evidence for a significant negative correlation between redshift and radio variability among
  bright FSRQs.
\end{abstract}

\section{Introduction}

Active galactic nuclei (AGN) are among the most energetic objects in the universe. In an AGN, a
variety of energetic phenomena are driven by a supermassive black hole (SMBH), fuelled by accretion
from its host galaxy. Many AGN produce relativistic jets: collimated structures
visible at parsec through kiloparsec scales. Relativistic beaming effects in the jets enhance the
intrinsically anisotropic appearance of the AGN, adding considerable complication to their
identification and to the interpretation of observations. Statistical studies involving large
samples are thus essential to the study of AGN, and careful attention must be paid to selection
effects~\citep[e.g.,][]{lister_statistical_1997}.

When the jet axis of an AGN is aligned very closely to our line of sight, Doppler-enhanced jet
emission dominates other emission processes and the source is classified as a blazar. Blazars emit
brightly across the entire electromagnetic spectrum, with a spectral energy distribution (SED)
characterized by two broad peaks. The lower peak, spanning the radio through the optical,
ultraviolet, or soft X-ray bands, is widely believed to be due to synchrotron emission from
relativistic electrons. The upper peak, which often extends to GeV and sometimes TeV gamma-ray
energies, is of less certain origin. It is most commonly ascribed to inverse Compton scattering of
low-energy seed photons by the synchrotron-emitting electrons in the jet, but other models remain
viable~\citep[e.g.,][]{boettcher_modeling_2007, boettcher_modeling_2012,
  dimitrakoudis_time-dependent_2012}. The source of the low-energy seed photons is
uncertain. Certainly some of the synchrotron radiation from within the jet is reprocessed and
emitted at higher energies, and models in which this process is predominant are known as synchrotron
self-Compton (SSC) models. The alternative is a source of seed photons external to the jet, such as
from the broad-line region, and models in which this is important are known as external Compton (EC)
models.

Blazars are customarily divided into two optical subclasses, the flat-spectrum radio quasars (FSRQs)
and the BL~Lac objects (BL~Lacs). Blazars with strong optical broad lines are classified as FSRQs,
while those with no lines or weak lines are classified as BL~Lacs. Although the BL~Lac and FSRQ
classes do seem to represent at least two distinct populations, there is evidence that this
classification scheme does not accurately reflect intrinsic physical differences between
sources. For example, the homogeneity of the BL~Lac class has been
questioned~\citep[e.g.,][]{marcha_optical_1996,anton_recognition_2005} and several nominal BL~Lacs
have been observed to meet the FSRQ definition in some epochs~\citep{shaw_fsrq_2012}. It remains
unclear how best to define physically meaningful classes. Several alternative classifications have
been proposed in recent
years~\citep[e.g.,][]{landt_physical_2004,ghisellini_transition_2011,meyer_blazar_2011,abdo_spectral_2010,giommi_simplified_2012}.

A promising and currently popular method is to classify blazars based on the frequency of the
synchrotron peak in the SED~\citep{padovani_giommi_1995,abdo_spectral_2010}. Blazars are divided
into low-, intermediate-, and high-synchrotron peaked (LSP, ISP, and HSP) sources, defined by
synchrotron peaks $\nupeak$ below $10^{14}~\mathrm{Hz}$, between $10^{14}$ and
$10^{15}~\mathrm{Hz}$, and above $10^{15}~\mathrm{Hz}$, respectively. FSRQs are almost exclusively
found to be LSP objects, while BL~Lac objects are found in all three spectral classes.  HSP BL~Lac
objects show clearly distinct gamma-ray and radio properties from
FSRQs~\citep[e.g.,][]{abdo_first_2010, abdo_spectral_2010, ackermann_second_2011,
  lister_gamma-ray_2011, linford_gamma-ray_2012}. HSP BL~Lacs display higher ratios of gamma-ray to
radio luminosity, lower radio core brightness temperatures, and appear to have lower jet Doppler
factors~\citep{lister_gamma-ray_2011}. There is also evidence that LSP BL~Lacs remain a distinct
class from FSRQs, though they appear to be sometimes misidentified~\citep{lister_gamma-ray_2011,
  linford_gamma-ray_2012}.

A connection between the radio and gamma-ray emission in blazars has long been suspected, and has
received renewed attention during the era of the \emph{Fermi} Large Area Telescope
(LAT)~\citep[e.g.][]{jorstad_multiepoch_2001,kovalev_relation_2009,
  ghirlanda_correlation_2010,mahony_high-frequency_2010, ackermann_radio/gamma_2011}. It is
straightforward to demonstrate a statistically significant correlation between observed
gamma-ray and radio luminosities or flux densities, as was done during the era of the EGRET
instrument~\citep[e.g.,][]{stecker_high-energy_1993,padovani_radio-loud_1993}. However, because of
distance effects, Malmquist bias, and the strong variability exhibited in blazars, such an apparent
correlation may or may not correspond to an interesting intrinsic
correlation~\citep[e.g.,][]{muecke_correlation_1997}. Recent studies using LAT data and concurrent
or nearly concurrent radio data provide strong evidence for an intrinsic
correlation~\citep{kovalev_relation_2009, ackermann_radio/gamma_2011}.  The data concurrency reduces
or eliminates spurious effects from variability, and Monte Carlo statistical methods demonstrate
that the correlation is not due to other
biases~\citep{ackermann_radio/gamma_2011,pavlidou_assessing_2012}. If the connection between radio
and gamma-ray emission is tight enough to produce correlated variability, cross-correlations between
light curves in the two bands could constrain the relative locations of the emission. A search for
statistically and physically significant time-domain correlations using our data set finds a few
examples, but longer time series are needed to rule out chance correlations in most
cases~(W.~Max-Moerbeck et al. submitted).

In this paper, we present results from our continuing investigation of the connection between the
gamma-ray loudness of a blazar and its radio variability.  In \citet{richards_blazars_2011}, we used
two years of data from our 15~GHz monitoring programme to find strong evidence that gamma-ray-loud
blazars in our radio-selected sample were more radio variable than their gamma-ray-quiet
counterparts. We also found significant differences in variability amplitudes between the
radio-selected FSRQ and BL~Lac populations, with the latter more radio variable.  The present paper
re-examines these findings using an additional two years of radio data, for a total of four
years. Our monitoring sample during this extended period now also contains a complete
gamma-ray-selected sample, the First LAT AGN Catalogue~\citep[1LAC;][]{abdo_first_2010}, which allows
us to compare the gamma-ray-selected FSRQ and BL~Lac populations.

\section{The Monitoring Programme}

Since late 2007, we have operated a fast-cadence 15~GHz radio blazar monitoring programme using the
40~m telescope at the Owens Valley Radio Observatory (OVRO).  Full details of the observing programme
are given in \citet{richards_blazars_2011} and \citet{jlr_thesis}. Total intensity flux density
measurements are performed using a combined Dicke-switching and beam-switching `double switching'
procedure to remove receiver gain fluctuation, atmospheric, and ground pick-up effects. Each flux
density is measured in about 70~s, including 32~s of on-source integration time. With a total system
temperature of about 55~K at zenith, including receiver, atmosphere, and cosmic microwave background
noise temperature contributions, this yields a typical measurement uncertainty of about 3~mJy, which
is measured from the scatter of the samples in each observation. Accurate pointing is achieved by
regularly peaking up on a relatively bright programme source. Residual pointing errors and other
sources contribute an additional 2~per cent uncertainty, added in quadrature with the measured error for
each measurement. The flux density scale is determined from regular observations of 3C~286
assuming the \citet{baars_absolute_1977} value of 3.44~Jy at 15.0~GHz, giving a 5~per cent overall scale
accuracy.

The monitoring sample began with the 1158 sources north of $-20\degr$ declination from the Candidate
Gamma-Ray Blazar Survey~\citep[CGRaBS,][]{healey_cgrabs:all-sky_2008}. The CGRaBS sources were
systematically selected to resemble AGN (mostly blazars) associated with EGRET gamma-ray detections
based on their radio spectral index, radio flux density, and X-ray flux.  In addition to the CGRaBS
sample, we have added the radio sources associated with \emph{Fermi}-LAT gamma-ray detections to our
programme. The sources included in the 1LAC catalogue compose a gamma-ray-selected sample. About 44~per cent of the sources in the 1LAC sample
are in the CGRaBS sample (263 of 599 at all declinations, 199 of 454 above
  $-20\degr$). These have been monitored since the inception of the programme. Some non-CGRaBS
sources were added after appearing in the LAT Bright AGN Sample (LBAS) based on the 3-month
\emph{Fermi}-LAT results~\citep{abdo_lbas_2009}. Any remaining 1LAC sources in our declination range
not already being monitored were added to the programme in March~2010.

In late 2011, we added sources in the Second LAT AGN Catalogue \citep[2LAC;][]{ackermann_second_2011}
to our observing programme. Although many were already in our programme, the observation period
described here does not include sufficient data to obtain robust results for some sources in
the complete 2LAC sample. We therefore use 1LAC as our gamma-ray
selected sample in the studies that follow. We do, however, use the 2LAC sample to identify the
subset of CGRaBS sources that are gamma-ray loud.

The CGRaBS and \emph{Fermi} samples differ substantially in the ratio of the FSRQs to BL~Lacs.  The
CGRaBS sample is dominated by FSRQs (70~per cent FSRQs vs. 11~per cent BL~Lacs), similar to the
ratio found by EGRET~\citep{hartman_3eg_1999,dermer_statistics_2007}. \emph{Fermi}-selected blazar samples are more evenly split, both for 1LAC (40~per cent
FSRQs vs. 49~per cent BL~Lacs) and 2LAC (45~per cent FSRQs vs 46~per cent BL~Lacs). This difference
results from a selection effect caused by the different spectral sensitivities of the EGRET and LAT
instruments, particularly to photons above about 1~GeV~\citep{abdo_fermi_2010, abdo_spectral_2010}.

To compare the radio variability of gamma-ray-loud HSP, ISP, and LSP objects, we use our 1LAC sample
and adopt the spectral classifications from the 1LAC catalogue~\citep{abdo_first_2010}. Our
454-source 1LAC sample comprises 99 HSP (22~per cent), 57 ISP (13~per cent), and 181 LSP (40~per
cent) sources, with 117 sources (26~per cent) unclassified\footnote{These percentages do not
    sum to exactly 100~per cent due to rounding.}. Of these, the BL~Lacs are predominantly HSP (98
sources, 44~per cent) with sizable fractions of ISP (53 sources, 24~per cent) and LSP (37 sources,
17~per cent), and 35 sources (16~per cent) unclassified. The FSRQs are almost exclusively LSP (134
sources, 73~per cent), with no HSP, one ISP, and 49~sources (27~per cent) unclassified.

For most objects we have adopted the redshift and optical classifications from the CGRaBS, 1LAC, or
2LAC catalogues~\citep{healey_cgrabs:all-sky_2008,abdo_first_2010,ackermann_second_2011}. Where these
disagreed, we adopted the 2LAC classification. In a few cases, we have adopted updated redshifts or
optical classifications based on an optical blazar spectroscopy campaign~\citep{shaw_fsrq_2012,
  shaw_bll_2013}. Our numerical values for \nupeak{} are those used for classification in the 2LAC
catalogue~\citep[B.~Lott personal comm.;][]{ackermann_second_2011}. Source names and associated
properties are listed in Table~\ref{tab:source}.

\begin{table*}
  \begin{minipage}{155mm}
    \caption{Source properties and results.}
    \label{tab:source}
    \begin{tabular}{@{}c c c c c c c c c}
      \hline
      CGRaBS Name & 1FGL Name & 2FGL Name & Opt. Class & SED Class & $z$ & $\im$ & $S_0$ & Faint? \\
      &&&&&&&(Jy)&\\
      \hline
      \mbox{\nodata} & \mbox{J0000.9-0745} & \mbox{J0000.9-0748} & BLL & ISP & \nodata & ${0.037\pm0.004}$ & ${0.169\pm0.001}$ &  \\
\mbox{J0001-1551} & \mbox{\nodata} & \mbox{\nodata} & FSRQ & \nodata & 2.044 & ${0.146\pm0.007}$ & ${0.236\pm0.002}$ &  \\
\mbox{J0001+1914} & \mbox{\nodata} & \mbox{\nodata} & FSRQ & \nodata & 3.100 & ${0.138\pm0.007}$ & ${0.260\pm0.002}$ &  \\
\mbox{J0003+2129} & \mbox{\nodata} & \mbox{\nodata} & \nodata & \nodata & 0.450 & ${0.113^{+0.008}_{-0.007}}$ & ${0.081\pm0.001}$ &  \\
\mbox{J0004-1148} & \mbox{\nodata} & \mbox{\nodata} & BLL & \nodata & \nodata & ${0.252\pm0.013}$ & ${0.618\pm0.010}$ &  \\
\mbox{J0004+2019} & \mbox{\nodata} & \mbox{\nodata} & BLL & \nodata & 0.677 & ${0.152\pm0.007}$ & ${0.355\pm0.003}$ &  \\
\mbox{J0004+4615} & \mbox{\nodata} & \mbox{\nodata} & FSRQ & \nodata & 1.810 & ${0.327^{+0.017}_{-0.016}}$ & ${0.174\pm0.004}$ &  \\
\mbox{J0005-1648} & \mbox{\nodata} & \mbox{\nodata} & \nodata & \nodata & \nodata & ${0.060\pm0.005}$ & ${0.160\pm0.001}$ &  \\

      \hline
    \end{tabular}

    \medskip Names of sources in the CGRaBS and 1LAC samples are specified in the first and second
    columns, respectively. CGRaBS sources with a corresponding entry in the `2FGL Name' column are
    considered gamma-ray loud. Sources that were dropped from variability analysis by our faintness
    criteria are indicated with an `F' in the `Faint?' column.
    (This is a table stub. The full table is available online in electronic form.)
  \end{minipage}
\end{table*}

\section{Results}
In this paper, we report results based on the four years of data collected between 2008
  January~1 and 2011 December~31, inclusive. The
light curve data used in these analyses are listed in Table~\ref{tab:data}; up-to-date data are
available from the programme website\footnote{\url{http://www.astro.caltech.edu/ovroblazars}}. The
data reduction pipeline has undergone several revisions to improve computational performance, better
reject unreliable measurements, and fix several programming errors.  This has resulted in slight
differences between the 2008--2009 data in this paper and that published in
\citet{richards_blazars_2011}.  None of these differences affect any of our conclusions.

\subsection{Variability Amplitude Trend}
\label{sec:variability_trend}
We begin by comparing the variability amplitudes found for each source from the four-year data set
with those reported for the two-year data set. To characterize the variability amplitude, we use the
intrinsic modulation index, \im{}, computed as described in \citet{richards_blazars_2011}. The
intrinsic modulation index is a maximum-likelihood estimate of the standard deviation of the source
flux density divided by its mean, ${\sigma_{S}/\left<S\right>}$. The uncertainty is also estimated,
accounting for measurement errors and varying numbers of measurements per source. As before, we
assume a Gaussian distribution of true flux density samples from each source. Although there are
several examples where this distribution is a poor fit to the data, our results are not sensitive to
this particular choice.  For each source, this method produces an intrinsic modulation index $\im$
and its uncertainty, reflecting the propagation of observational uncertainties into $\im$.  We also
compute a maximum-likelihood mean flux density, $S_0$, and its uncertainty for each source. Sources
for which more than 10~per cent of the measured flux densities were non-detections at $2\sigma$, for
which the mean flux density was negative, or for which the ratio of the mean flux density to its
error was less than two were deemed too faint for variability analysis and excluded. These criteria
excluded 7 of the 1158 CGRaBS sources and 76 of the 454 1LAC sources in our samples. Mean fluxes and
intrinsic modulation indices are listed in Table~\ref{tab:source}.

Intuitively, we expect that additional data will tend to increase the variability amplitude on
average.  Many sources are observed to switch between periods of relatively steady quiescence and
periods of active variability. The addition of a period of steady flux to a source with
a history of strong variability will reduce its intrinsic modulation index only slightly because the
amplitude of variability is dominated by the largest excursions. On the other hand, a source that
has only been observed in a weakly variable state will see a large increase in its intrinsic
modulation index if it begins to vary strongly. With additional observations, we expect some sources
with weak or no apparent variability in the first two years will `turn on' and exhibit significant
increases in variability.

\begin{figure}
  \centering
  \includegraphics[width=\columnwidth]{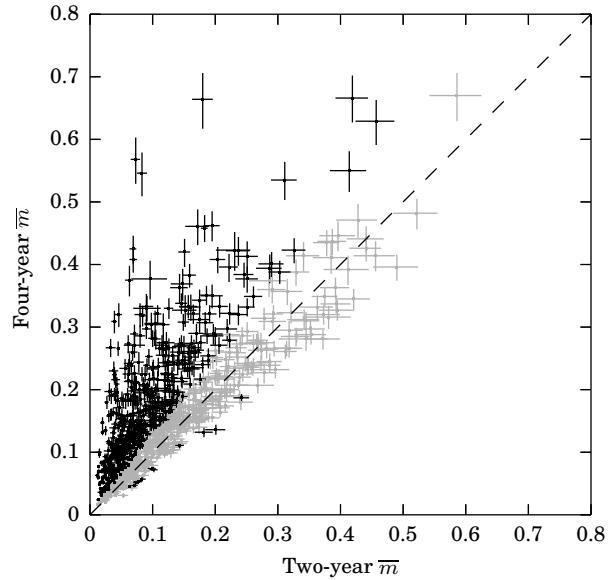}
  \caption{Scatter plot of four-year versus two-year modulation indices for 1134 CGRaBS sources. The
    dashed line shows the 1:1 relationship for reference.  Sources for which the difference in
    intrinsic modulation index is less than $3\sigma$ are plotted in grey.}
  \label{fig:modindex_scatter}
\end{figure}

The data confirm our expectations. Between the two-year and four-year data, the change in the
intrinsic modulation index for each source is an increase of 0.044 (mean) or 0.029 (median).  In
Fig.~\ref{fig:modindex_scatter} we plot the four-year \im{} values against the two-year \im{}
values for the 1134 CGRaBS sources with measured \im{} in both data sets. The
black points represent the 598 of 1134 sources that exhibited more than a
$3\sigma$ change. Our calibration sources are consistent with no change, except for \mbox{3C
    274} (which went from \valunc{0.009}{0.001} to \valunc{0.017}{0.001}) and \mbox{3C 286} (which
  went from \valunc{0.006}{0.001} to \valunc{0.011}{0.001}). The former is known to vary slowly, so
  this may result in part from real variation. For both sources, the intrinsic modulation indices in
  both data sets are below our cutoff for inclusion in population comparisons (0.02). For one source,
  \mbox{J1154+1225}, we found a single very large outlier in the two-year light curve that resulted
  in an erroneously high intrinsic modulation index in \citet{richards_blazars_2011}. When we
  reanalyse the light curve with this outlier removed, we obtain an upper limit slightly below the
  four-year intrinsic modulation index. Because upper limits are not plotted in
  Fig.~\ref{fig:modindex_scatter}, this source is not included.

  This systematic increase in the variability index suggests that the two year interval was
  insufficiently long to capture the full range of behaviors in many CGRaBS sources: variability on
  time-scales longer than two years is apparent.  This is not surprising.  Based on more than
  25~years of monitoring at the University of Michigan Radio Astronomy Observatory and the
  Mets\"ahovi Radio Observatory in Finland, typical flaring time-scales of 4--6~years were found,
  sometimes with evidence for changes on time-scales of 10~years or longer, and typical flare
  durations of 2.5~years were reported at 22~and 37~GHz~\citep{hovatta_statistical_2007,
    hovatta_long-term_2008}.  Long-time-scale variability is also found
  in the gamma-ray band. For example, variability on few-year time-scales led to lower ratios
  between the peak and mean gamma-ray fluxes detected by the LAT during its first 11~months than
  were found by EGRET during its 4.5~year lifetime~\citep{abdo_first_2010}. Thus, as our monitoring
  programme continues, additional data will likely continue to increase the intrinsic modulation
  index for some sources. However, the affected sources will be distributed randomly among the
  subpopulations we use in our studies, so this trend will not create false apparent correlations.

\begin{table}
  \caption{15~GHz light curve data.}
  \label{tab:data}
  \begin{tabular}{@{} c c c c}
    \hline
    CGRaBS Name & 1FGL Name & MJD & Flux Density \\
    &&(days)&(Jy)\\
    \hline
    \mbox{\nodata} & \mbox{J0000.9-0745} & 55310.785035 & \mbox{$0.166\pm0.006$} \\
\mbox{} & \mbox{} & 55320.769120 & \mbox{$0.154\pm0.008$} \\
\mbox{} & \mbox{} & 55324.757593 & \mbox{$0.158\pm0.004$} \\
\mbox{} & \mbox{} & 55331.738160 & \mbox{$0.161\pm0.006$} \\
\mbox{} & \mbox{} & 55337.715532 & \mbox{$0.173\pm0.007$} \\
\mbox{} & \mbox{} & 55340.706921 & \mbox{$0.175\pm0.006$} \\
\mbox{} & \mbox{} & 55349.623113 & \mbox{$0.165\pm0.005$} \\
\mbox{} & \mbox{} & 55352.608426 & \mbox{$0.164\pm0.005$} \\
\mbox{} & \mbox{} & 55355.600590 & \mbox{$0.164\pm0.008$} \\
\mbox{} & \mbox{} & 55361.581215 & \mbox{$0.168\pm0.011$} \\

    \hline
  \end{tabular}

  \medskip
  This is a table stub. The full table is available online in electronic form.
\end{table}

\subsection{Outlier Contamination}
\label{sec:outliers}

\begin{figure}
  \centering
  \includegraphics[width=\columnwidth]{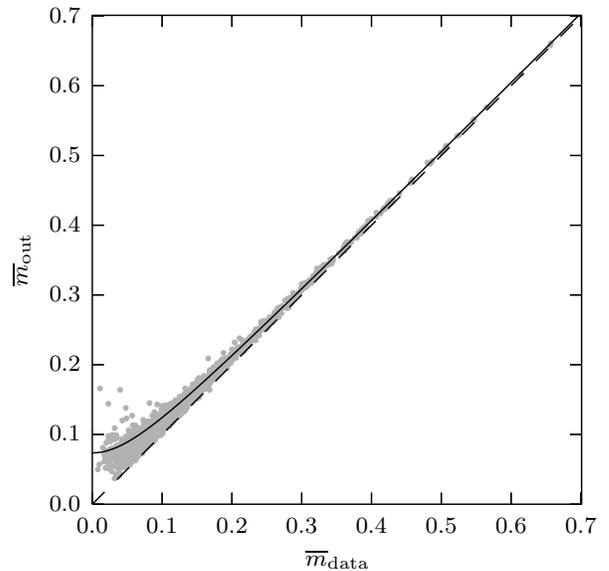}
  \caption{Grey points show modulation index data computed with the addition of an extreme outlier
    data point plotted against the modulation index for the same source calculated from the actual
    data. The dashed line shows the ideal $y=x$ line. The solid line shows the effect of adding
    0.066 in quadrature with the measured modulation index.}
  \label{fig:m_out_vs_m_real}
\end{figure}

Although our automated data quality filters reject most unreliable observations, our flux density
light curves still contain some outlier data points. These are mostly attributable to poor observing
conditions or pointing offset measurement failures, and do not represent actual astronomical source
variations. These outliers will artificially increase the intrinsic modulation index we compute for
an affected source, so we must account for their effect. We cannot simply delete these points from
the light curves on the basis that the flux density value differs from neighboring measurements:
blazars are strongly variable objects, and such deletion would bias our results toward our
preconception of `reasonable' variability. To avoid such biases, we instead quantify how the
presence of outliers affects our calculated intrinsic modulation indices.

The most common extreme outliers are near-zero flux density values reported for normally bright
sources.  These are probably due to mis-pointing. High outliers do occur occasionally, though less
frequently because they are more effectively removed by our data quality filters. To measure the
effect of outliers, we computed the intrinsic modulation indices for each source with the addition
of a single `false outlier' flux density value that was twice the average error above zero. High
outliers are rarely more than twice the true flux density of a source, so their impact on the
intrinsic modulation index is similar to that of the near-zero outliers. We performed this test
using modulation indices computed from the first 3.5~years of data for each source. The result is
approximately described by a quadrature increase of the intrinsic modulation index by 0.066, as
shown in Fig.~\ref{fig:m_out_vs_m_real}. 

We estimate that about 8~per cent of our sources are affected by such
outliers~\citep{jlr_thesis}. Because of this low incidence, sources affected by multiple outliers
are rare ($\lesssim\!2$~per cent). The impact of a second outlier is smaller than that of the first
because the intrinsic modulation index is dominated by the largest excursion from the mean. In
\citet{jlr_thesis}, addition of a second simulated outlier was found to increase the intrinsic
modulation by a mean of only 0.001. We therefore use the effect of a single outlier to estimate the
impact on our modulation index results. The incidence of outlier data points should not be
correlated with physical properties of the sources, so the net effect of outliers is to increase the
average variability of a population, slightly reducing our sensitivity to differences between
subpopulations.

\subsection{Trends with SED Peak Frequency}
\begin{figure}
  \centering
  \includegraphics[width=\columnwidth]{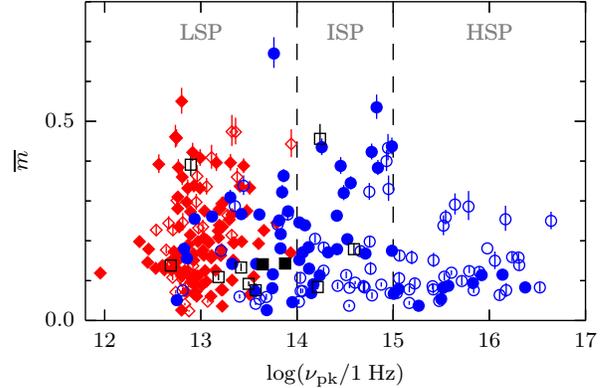}
  \caption{Intrinsic modulation index \im{} versus \nupeak{} for the 258 sources in the 1LAC sample
    with both measured 15~GHz modulation indices in this work and \nupeak{} values in
    \citet{ackermann_second_2011}. BL~Lacs are plotted as blue circles, FSRQs as red diamonds, and
    other classifications as black squares. CGRaBS sources are plotted as filled symbols.}
  \label{fig:im_vs_nupeak}
\end{figure}

\begin{figure}
  \centering
  \includegraphics[width=\columnwidth]{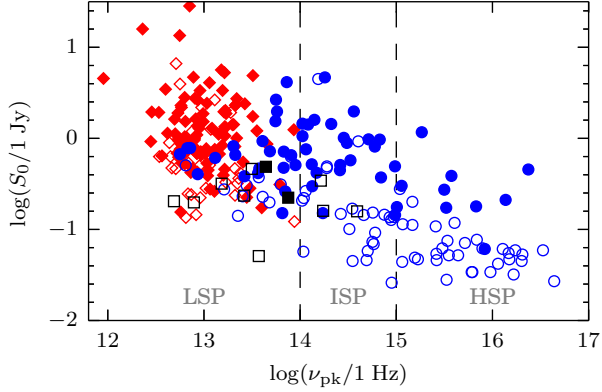}
  \caption{Maximum likelihood mean flux density, $S_0$ versus \nupeak{} for the 248 BL~Lacs and
    FSRQs in the 1LAC sample with both 15~GHz mean flux density values in this work and \nupeak{}
    values in \citet{ackermann_second_2011}. BL~Lacs are plotted as blue circles, FSRQs as red diamonds, and
    other classifications as black squares. CGRaBS sources are plotted as filled symbols.}
  \label{fig:flux_vs_nupeak}
\end{figure}

When the intrinsic modulation index is plotted against \nupeak{} in Fig.~\ref{fig:im_vs_nupeak}, the
result appears familiar. Its form strongly resembles equivalent plots of radio core brightness
temperatures and fractional polarizations~\citep{lister_gamma-ray_2011}, gamma-ray variability
indices~\citep{ackermann_second_2011}, and optical intrinsic modulation
indices~\citep{hovatta_connection_2013}. Fig.~\ref{fig:im_vs_nupeak} differs from these in that
there is a rather abrupt decrease in the apparent upper envelope above $10^{15}~\mathrm{Hz}$, where
the others show a smoother transition in the ISP region (for the gamma-ray variability indices and
optical modulation indices) or a step decrease at a slightly lower \nupeak{} value (nearer to
$10^{14.5}~\mathrm{Hz}$ for the radio core brightness temperatures and
fractional polarizations). We note that removal of only a few high-$\im$,
high-\nupeak{} ISP sources would eliminate the abrupt decrease in our plot, so this could be the
result of incorrect \nupeak{} values. Using the two-sample Kolmogorov--Smirnov test to compare the
$\im$ values for the three classes, we find that neither the LSP nor the ISP sample is consistent
with the null hypothesis of being drawn from the same distribution as the HSP sample
(${p\approx7\times10^{-4}}$ and ${p\approx0.002}$). For the LSP and ISP samples, we cannot reject
the null hypothesis (${p\approx0.6}$). As shown in Fig.~\ref{fig:flux_vs_nupeak}, sources with lower
\nupeak{} values tend to have higher mean flux densities. This relationship matches that found for
parsec-scale 15~GHz flux densities by \citet{lister_gamma-ray_2011}. Because HSP sources are found
at lower flux densities, we measure a value for \im{} for proportionally fewer of these sources. We
have not accounted for upper limits in this analysis, but doing so would likely increase the
difference between the HSPs and the other classes.

\begin{figure}
  \centering
  \includegraphics[width=\columnwidth]{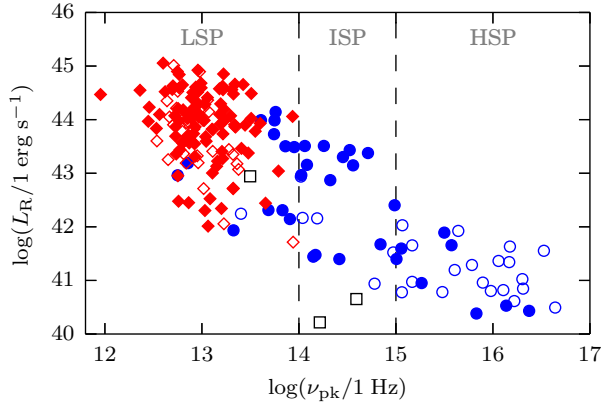}
  \caption{Isotropic radio luminosity, $L_{\mathrm{R}}$ versus \nupeak{} for the 186 BL~Lacs and
    FSRQs in the 1LAC sample with 15~GHz mean flux density values in this work, known redshifts, and
    \nupeak{} values in \citet{ackermann_second_2011}. BL~Lacs are plotted as blue circles, FSRQs as
    red diamonds, and other classifications as black squares. CGRaBS sources are plotted as filled
    symbols.}
  \label{fig:lum_vs_nupeak}
\end{figure}

In Fig.~\ref{fig:lum_vs_nupeak}, we plot the isotropic radio luminosity computed as
\begin{equation}
  L_{\mathrm{R}}=\frac{4\pi D_{\mathrm{L}}^2 \Delta\nu S_0}{1+z},
\end{equation}
where $D_{\mathrm{L}}$ is the luminosity distance (assuming a flat $\Lambda$CDM cosmology with
$\Omega_{\mathrm{\Lambda}}{=}0.726$ and $H_0{=}70.5~\mathrm{km\ s^{-1}\ Mpc^{-1}}$),
$\Delta\nu{=}3~\mathrm{GHz}$ is the bandwidth, $S_0$ is the maximum-likelihood mean flux density, and $z$
is the redshift. Both \nupeak{} and $L_{\mathrm{R}}$ are strongly correlated with redshift (for the
entire sample, we find Kendall's ${\tau=-0.44}$ and ${\tau=0.70}$, respectively, both with
${p<10^{-16}}$)\footnote{Kendall's tau correlation and partial correlation coefficients were
  computed in R using method=`kendall' with the native cor.test() function and the pcor.test()
  function from the \textsc{ppcor} package~\citep{r-package,ppcor-package}.}. To account for this mutual
correlation, we compute the partial correlation coefficient for ${L_\mathrm{R}}$ versus \nupeak{}
while controlling for $z$. We find a highly significant correlation for BL~Lacs (${\tau=-0.48}$ with
${p<10^{-7}}$) and no significant correlation for FSRQs (${\tau=-0.10}$ with ${p<0.10}$). Partial
correlations using flux densities instead of isotropic luminosities give the same results. Although
the luminosity correlation is not formally significant for FSRQs, we note that they do coincide with
the low-$\nupeak$ end of the BL~Lac trend. A correlation among FSRQs may simply be difficult to
detect because of the narrow range of \nupeak{} values found in these sources.

\section{Population Comparisons}

To compare the variability amplitude between source populations in our sample, we apply a likelihood
maximization method that requires we specify a parent distribution for the modulation indices. An
exponential distribution, $f(m)=m_{0}^{-1}\exp(-m/m_0)$, is a qualitatively reasonable fit to the
observed distribution of modulation indices in our sample~\citep{richards_blazars_2011}. This
distribution is a monoparametric distribution characterized by its mean, $m_0$. Because of this,
numerical integrations are required in only one dimension, making it convenient for our likelihood
analyses.

As in \citet{richards_blazars_2011}, we estimate $m_0$ for a population via likelihood maximization
using an analysis that avoids bias by excluding regions of the $(S_0,\im)$ parameter space where we
do not adequately measure \im{}. For a source $i$, the likelihood of observing a
modulation index $\im_i$ with a Gaussian uncertainty $\sigma_i$ drawn from an exponential
distribution with mean $m_0$ is
\begin{eqnarray}
\ell[\im_i]&=& \frac{1}{2m_0}
\exp\left[-\frac{\im_i}{m_0}\left(1-\frac{\sigma_i^2}{2m_0\im_i}\right)\right]\times \nonumber \\
&& \left\{1+{\rm erf}\left[ \frac{\im_i}{\sigma_i\sqrt{2}} \left(1-\frac{\sigma_i^2}{m_0\im_i}\right)\right] \right\}.
\end{eqnarray}
If we have excluded data where $\im_i<\im_{\mathrm{L}}$ for some lower limit $\im_{\mathrm{L}}$,
then we must correct this calculation to reflect this. This gives
\begin{equation}
\ell_{\mathrm{cuts}}[\im_i, \im_{\mathrm{L}}] = \frac{H(\im_i-\im_{\mathrm{L}})\ell[\im_i]}{\int _{\im_{\mathrm{L}}}^\infty \ell[\im_i^{\prime}] d\im_i^{\prime}}\,,
\end{equation}
where $H$ is the Heaviside step function.  In the population studies described here, we consider
only sources for which we found a mean 15~GHz flux density ${S_0>60~\mathrm{mJy}}$ and for which
${\im>0.06}$ (if ${S_0<400~\mathrm{mJy}}$) or ${\im>0.02}$ (if ${S_0\geq400~\mathrm{mJy}}$). This
corresponds to two separate cuts, one at ${\im_{\mathrm{L}}=0.06}$ and the other at
${\im_{\mathrm{U}}=0.02}$. With ${S_0=400~\mathrm{mJy}}$ as a threshold, we divide the sources into
two groups, with $N_{\mathrm{L}}$ and $N_{\mathrm{U}}$ members. Of the 1151 CGRaBS sources bright
enough for variability analysis, we drop 9 sources because ${S_0<60~\mathrm{mJy}}$, 95
because ${\im<\im_{\mathrm{L}}}$, and 4 because ${\im<\im_{\mathrm{U}}}$. Of the 378 1LAC sources
bright enough for variability analysis, we drop 36 because ${S_0<60~\mathrm{mJy}}$, 18
because $\im<\im_{\mathrm{L}}$, and 2 because ${\im<\im_{\mathrm{U}}}$. We treat the groups as
separate experiments and the total likelihood to observe $m_0$ is simply the product of the
individual likelihoods,
\begin{equation}\label{likelihood}
\mathcal{L} (m_0)= \prod_{i=1}^{N_{\mathrm{L}}} \ell_{\rm cuts}[\im_i, \im_{\mathrm{L}}]\prod_{i=1}^{N_{\mathrm{U}}} \ell_{\rm cuts}[\im_i, \im_{\mathrm{U}}]\,.
\end{equation}
Results are tabulated in Table~\ref{tab:cgrabs_modindices} for the
CGRaBS sample and Table~\ref{tab:1lac_modindices} for the 1LAC
sample. For each subpopulation, the reported value is the most likely
value in the distribution. The upper and lower uncertainties
correspond to the equal-likelihood points between which the integrated
area under the distribution contains 68.26~per cent of the total.  The
likelihood distribution of the difference in the mean intrinsic
modulation index between two subpopulations, $\Delta m_0$, is given by
the cross-correlation of the individual likelihood distributions.  The
results of these comparisons are listed in
Table~\ref{tab:population_results}.

\begin{table}
  \caption{CGRaBS subpopulation intrinsic modulation indices.}
  \label{tab:cgrabs_modindices}
  \begin{tabular}{@{}c c}
    \hline
    Subpopulation & $m_{0}$ \\
    \hline
    gamma-ray loud& \valunca{0.175}{0.011}{0.012} \\[1.5ex]
    gamma-ray quiet & \valunca{0.099}{0.003}{0.004}\\[1.5ex]
    BL~Lac                          & \valunca{0.163}{0.014}{0.016}    \\[1.5ex]
    FSRQ                             & \valunc{0.112}{0.004}   \\[1.5ex]
    FSRQ ($z<1$)                            & \valunca{0.123}{0.007}{0.008} \\[1.5ex]
    FSRQ ($z\geq1$)                        & \valunc{0.106}{0.005}         \\[1.5ex]
    \hline
  \end{tabular}

  \medskip
  A source is included in the gamma-ray loud subpopulation if it has a clean association in the
  2LAC catalogue~\citep{ackermann_second_2011}.
\end{table}

\begin{table}
  \caption{1LAC subpopulation intrinsic modulation indices.}
  \label{tab:1lac_modindices}
  \begin{tabular}{@{}c c}
    \hline
    Subpopulation & $m_{0}$ \\
    \hline
    BL~Lac                          & \valunca{0.150}{0.014}{0.015} \\[1.5ex]
    FSRQ                             & \valunca{0.181}{0.013}{0.014} \\[1.5ex]
    HSP                             & \valunca{0.036}{0.008}{0.010} \\[1.5ex]
    ISP                             & \valunca{0.175}{0.025}{0.031} \\[1.5ex]
    LSP                              & \valunca{0.177}{0.013}{0.014} \\[1.5ex]
    HSP BL~Lac                      & \valunca{0.036}{0.008}{0.010} \\[1.5ex]
    ISP BL~Lac                      & \valunca{0.178}{0.026}{0.033} \\[1.5ex]
    LSP BL~Lac                       & \valunca{0.155}{0.027}{0.036}\\[1.5ex]
    \hline
  \end{tabular}
\end{table}

\begin{table*}
  \begin{minipage}{90mm}
    \caption{Population variability comparison results.}
    \label{tab:population_results}
    \begin{tabular}{@{}c c c c c c c}
      \hline
      Parent Pop. & Subpop. A & Subpop. B & $\Delta m_0$ & Signif.\\
      \hline
      CGRaBS & gamma-ray loud & gamma-ray quiet & \valunca{0.075}{0.012}{0.013} & $6\sigma$ \\[1.5ex]
      CGRaBS & BL~Lac                          & FSRQ                             & \valunca{0.050}{0.015}{0.017} & $4\sigma$ \\[1.5ex]
      1LAC   & BL~Lac                          & FSRQ                             & \valunc{-0.031}{0.020}        & $<\!2\sigma$ \\[1.5ex]
      BL~Lac & CGRaBS                          & 1LAC                             & \valunc{0.013}{0.021}         & $<\!1\sigma$\\[1.5ex]
      FSRQ   & CGRaBS                          & 1LAC                             & \valunca{-0.068}{0.015}{0.014}& $6\sigma$ \\[1.5ex]
      1LAC   & HSP                             & ISP                              & \valunca{-0.136}{0.032}{0.027}& $5\sigma$ \\[1.5ex]
      1LAC   & HSP                             & LSP                              & \valunc{-0.139}{0.017}        & $5\sigma$ \\[1.5ex]
      1LAC   & ISP                             & LSP                              & \valunca{-0.002}{0.029}{0.033}& $<\!1\sigma$\\[1.5ex]
      1LAC   & HSP BL~Lac                      & ISP BL~Lac                       & \valunca{-0.139}{0.034}{0.028}& $4\sigma$ \\[1.5ex]
      1LAC   & HSP BL~Lac                      & LSP BL~Lac                       & \valunca{-0.116}{0.037}{0.029}& $4\sigma$ \\[1.5ex]
      1LAC   & ISP BL~Lac                      & LSP BL~Lac                       & \valunca{0.022}{0.045}{0.044}& $<\!1\sigma$\\[1.5ex]
      CGRaBS & FSRQ ($z\geq1$)                 & FSRQ ($z<1$)                     & \valunc{-0.018}{0.009}        & $<\!2\sigma$\\
      \hline
    \end{tabular}

    \medskip
    The $\Delta m_0$ column tabulates the most likely value of $m_{0,\mathrm{A}}-m_{0,\mathrm{B}}$.
    A source is included in the gamma-ray loud subpopulation if it has a clean association in the
    2LAC catalogue~\citep{ackermann_second_2011}.
  \end{minipage}
\end{table*}

\subsection{Gamma-ray Loudness}
\begin{figure}
  \centering
  \includegraphics[width=\columnwidth]{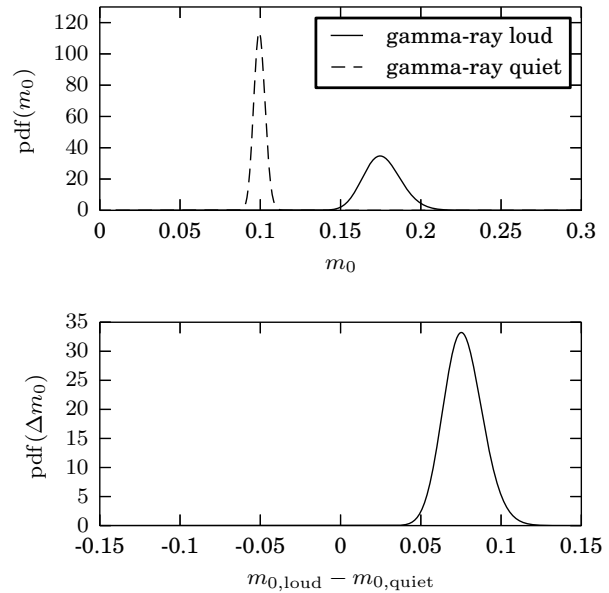}
  \caption{Top: Likelihood distributions for $m_0$ for CGRaBS sources
    that are gamma-ray loud (solid line) and gamma-ray quiet (dashed
    line). The most likely value of the mean modulation index for each
    distribution and for the difference between the two are listed in
    Tables~\ref{tab:cgrabs_modindices}
    and~\ref{tab:population_results}. A source is considered gamma-ray
    loud if it is has a clean association in the \emph{Fermi} 2LAC
    catalogue~\citep{ackermann_second_2011}.  Bottom: Likelihood
    distribution of the difference between the mean modulation
    indices. This distribution is inconsistent with a zero mean with
    about $6\sigma$ significance.}
  \label{fig:cgrabs_loud_vs_quiet}
\end{figure}

Using our population comparison method, we now investigate whether the connection between gamma-ray
emission and radio variability we reported in \citet{richards_blazars_2011} persists in our longer
data set.  We define a CGRaBS source to be gamma-ray loud if it has a `clean' association with a
LAT gamma-ray source in the 2LAC catalogue. The likelihood distributions for the
  gamma-ray-loud and quiet subpopulations are shown in
Fig.~\ref{fig:cgrabs_loud_vs_quiet}.  To evaluate the significance of the separation between the
distributions, we compute the likelihood distribution for the difference between the true population
means, also shown in Fig.~\ref{fig:cgrabs_loud_vs_quiet}.  The two distributions are not
consistent with each other at the $6\sigma$ confidence level with a most likely difference of
7.5~per cent. As before, gamma-ray-loud sources exhibit higher variability amplitudes than do
gamma-ray-quiet sources.

\subsection{Optical Classification}

\begin{figure}
  \centering
  \includegraphics[width=\columnwidth]{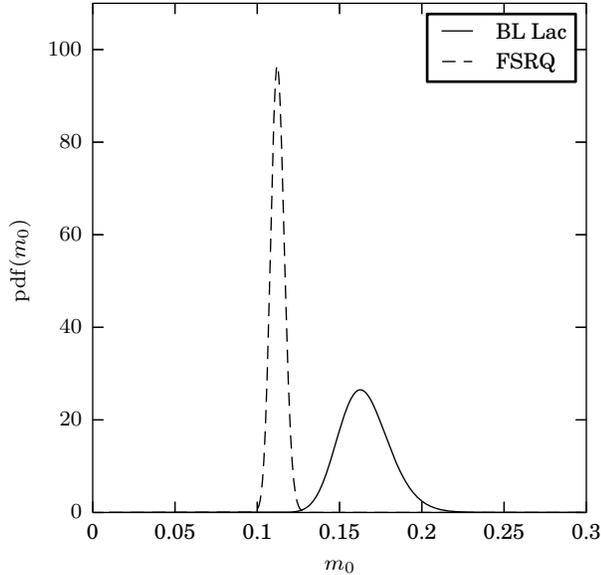}
  \caption{Likelihood distributions for $m_0$ for CGRaBS BL~Lacs
    (solid line) and FSRQs (dashed line). The most likely values for the
    mean modulation index for each distribution and for the difference
    between the two are listed in Tables~\ref{tab:cgrabs_modindices}
    and~\ref{tab:population_results}. The two distributions \emph{are
      not} consistent with having the same mean modulation index with
    about $4\sigma$ significance.}
  \label{fig:cgrabs_bll_vs_fsrq}
\end{figure}

We next examine radio variability amplitude as a function of optical classification, comparing the
BL~Lac and FSRQ subpopulations. In Fig.~\ref{fig:cgrabs_bll_vs_fsrq}, we show the likelihood
distributions for $m_0$ for sources in the CGRaBS sample. The CGRaBS BL~Lacs are found to be more
variable at a significance of about $4\sigma$, with a most likely difference of
\valunca{0.050}{0.015}{0.017}. This is consistent with the results previously reported for the
two-year data set~\citep{richards_blazars_2011}.

\begin{figure}
  \centering
  \includegraphics[width=\columnwidth]{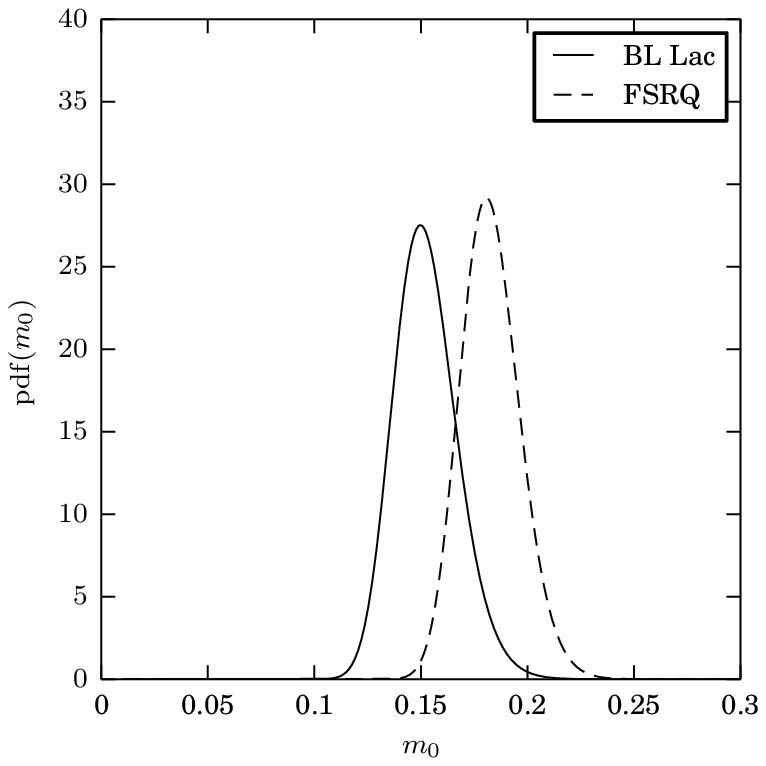}
  \caption{Likelihood distributions for $m_0$ for 1LAC BL~Lacs (solid
    line) and FSRQs (dashed line). The most likely values for the mean
    modulation index for each distribution and for the difference
    between the two are listed in Tables~\ref{tab:1lac_modindices}
    and~\ref{tab:population_results}. The two distributions \emph{are}
    consistent with having the same mean modulation index at the
    $2\sigma$ level.}
  \label{fig:1lac_bll_vs_fsrq}
\end{figure}

The situation is quite different for the 1LAC sample. Likelihood distributions for $m_0$ for the
BL~Lac and FSRQ subpopulations of the 1LAC sample are shown in Fig.~\ref{fig:1lac_bll_vs_fsrq}. We
find only weak evidence for a difference in $m_0$ between the BL~Lac and FSRQ subpopulations (most
likely difference \valunc{0.031}{0.020}, corresponding to $<\!2\sigma$ significance). Although the
difference is not statistically significant, we note that the mean modulation index for BL~Lacs is
formally lower than that for FSRQs, while for the CGRaBS sample we find the variability among FSRQs
to be significantly higher.

\begin{figure}
  \centering
  \includegraphics[width=\columnwidth]{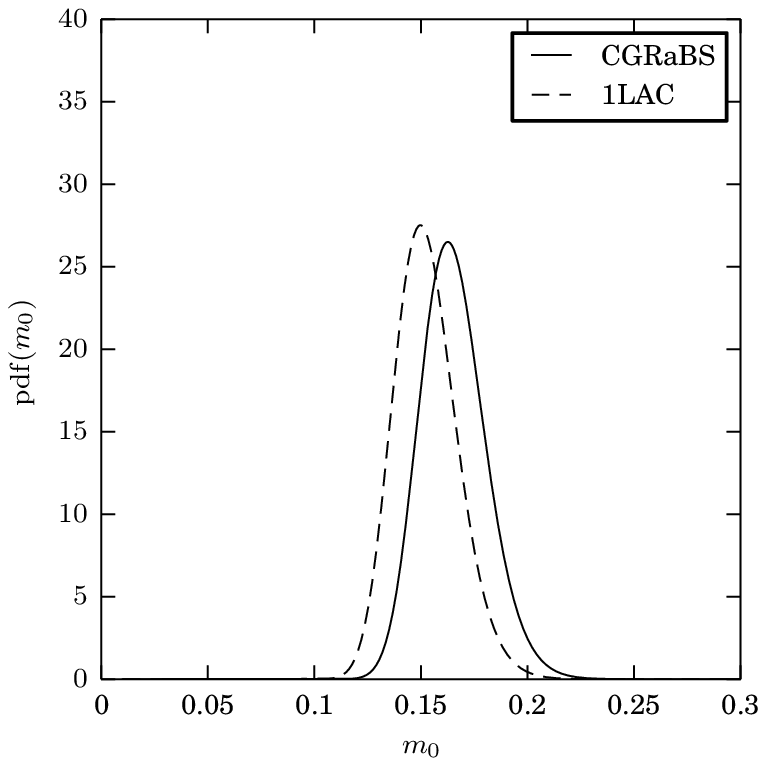}
  \caption{Likelihood distributions for $m_0$ for BL~Lacs in CGRaBS
    (solid line) and 1LAC (dashed line). The most likely values for the
    mean modulation index for each distribution and for the difference
    between the two are listed in
    Tables~\ref{tab:cgrabs_modindices}--\ref{tab:population_results}. The
    two distributions \emph{are} consistent with having the same mean
    modulation index at the $1\sigma$ level. }
  \label{fig:bll_1lac_vs_cgrabs}
\end{figure}

\begin{figure}
  \centering
  \includegraphics[width=\columnwidth]{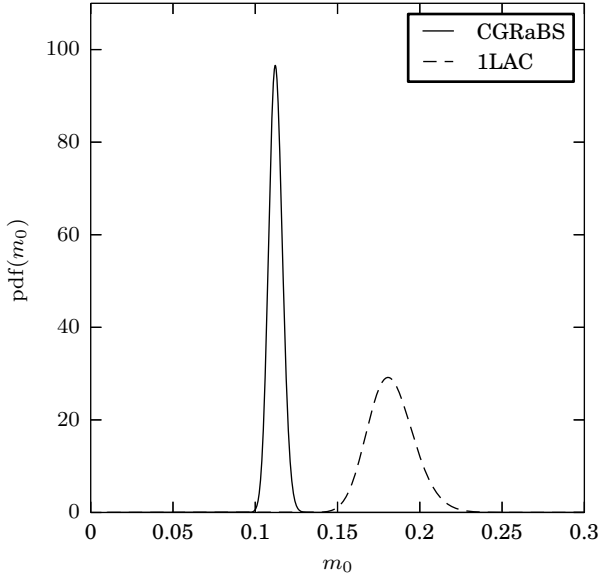}
  \caption{Likelihood distributions for $m_0$ for FSRQs in CGRaBS
    (solid line) and 1LAC (dashed line). The most likely values for the
    mean modulation index for each distribution and for the difference
    between the two are listed in
    Tables~\ref{tab:cgrabs_modindices}--\ref{tab:population_results}. The
    two distributions \emph{are not} consistent with having the same
    mean modulation index with about $6\sigma$ significance.}
  \label{fig:fsrq_1lac_vs_cgrabs}
\end{figure}

As a further test, we compare the variability amplitudes of the CGRaBS and 1LAC samples in
Fig.~\ref{fig:bll_1lac_vs_cgrabs} (for BL~Lacs) and Fig.~\ref{fig:fsrq_1lac_vs_cgrabs}
(for FSRQs). Note that the individual likelihood distributions shown in
Figs.~\ref{fig:bll_1lac_vs_cgrabs} and~\ref{fig:fsrq_1lac_vs_cgrabs} are the same as those in
Figs.~\ref{fig:cgrabs_bll_vs_fsrq} and~\ref{fig:1lac_bll_vs_fsrq}, but are plotted in different
pairs.  We find no evidence that the BL~Lacs in the 1LAC and CGRaBS samples differ in variability
amplitude, with a most-likely difference of \valunc{0.013}{0.021}.  In contrast, 1LAC FSRQs are more
variable than the CGRaBS FSRQs with a most-likely difference of \valunca{0.068}{0.014}{0.015}, a
difference significant at the $6\sigma$ level.  Thus, the difference in radio variability between
gamma-ray-loud and gamma-ray-quiet CGRaBS blazars reflects a large variability difference between
FSRQs in the mostly radio-selected CGRaBS sample and the gamma-ray-selected sample. BL~Lacs in
either sample exhibit similar radio variability.

\subsection{Spectral Classification}

\begin{figure}
  \centering
  \includegraphics[width=\columnwidth]{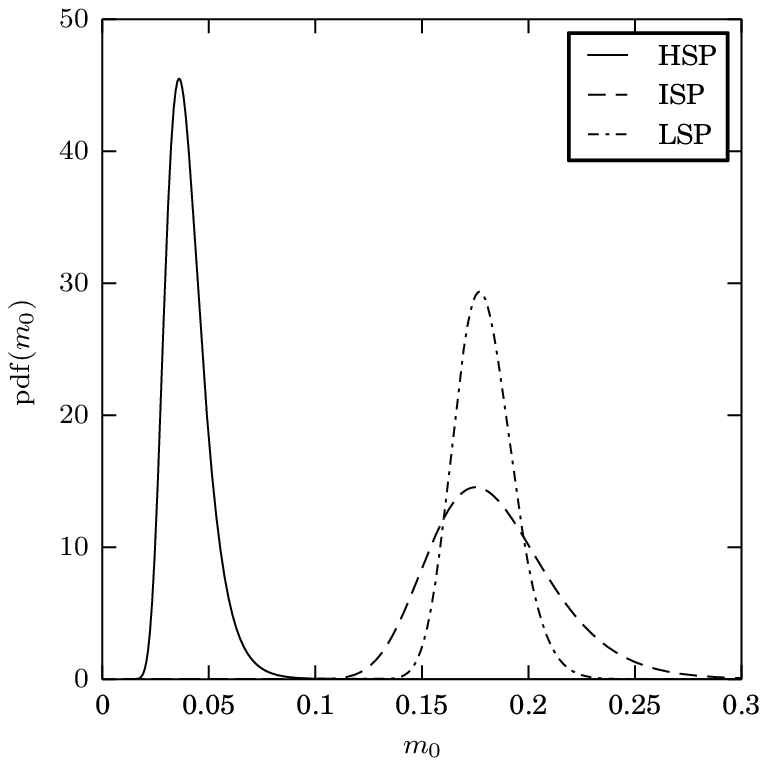}
  \caption{Likelihood distributions for $m_0$ for 1LAC HSP sources
    (solid line), ISP sources (dashed line), and LSP sources
    (dot-dashed line). The most likely values for the mean modulation
    index for each distribution and for the differences between the
    pairs are listed in Tables~\ref{tab:1lac_modindices}
    and~\ref{tab:population_results}. The ISP and LSP distributions
    \emph{are} consistent with having the same mean modulation index
    at the $1\sigma$ level. The HSP distribution \emph{is not}
    consistent with either of the others with about $5\sigma$
    significance.}
  \label{fig:hsp_isp_lsp}
\end{figure}

Comparing the HSP, ISP, and LSP populations in the 1LAC sample, we find that the HSP population is
less variable than either of the others, while the LSP and ISP populations are not distinguishable.
The likelihood distributions are plotted in Fig.~\ref{fig:hsp_isp_lsp}. The most likely difference
between HSP and LSP populations is \valunc{0.139}{0.017}
and between the HSP and ISP populations is
\valunca{0.136}{0.027}{0.032}, both significant at about
the $5\sigma$ level. Between the ISP and LSP populations, the difference is only
\valunca{0.002}{0.033}{0.029}.

\begin{figure}
  \centering
  \includegraphics[width=\columnwidth]{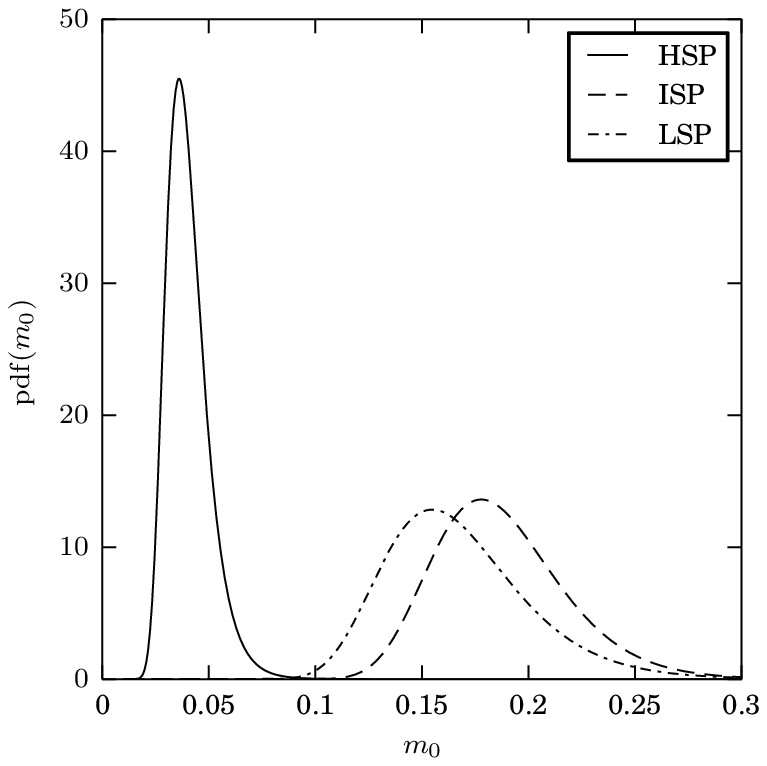}
  \caption{Likelihood distributions for $m_0$ for 1LAC BL~Lac sources
    classified as HSP, (solid line), ISP (dashed line), and LSP
    (dot-dashed line). The most likely values for the mean modulation
    index for each distribution and for the differences between the
    pairs are listed in Tables~\ref{tab:1lac_modindices}
    and~\ref{tab:population_results}. The ISP and LSP distributions
    \emph{are} consistent with having the same mean modulation index
    at the $1\sigma$ level. The HSP distribution \emph{is not}
    consistent with either of the others with about $4\sigma$
    significance.}
  \label{fig:bll_hsp_isp_lsp}
\end{figure}

In Fig.~\ref{fig:bll_hsp_isp_lsp}, the modulation index likelihood distributions are plotted for
the HSP, ISP, and LSP BL~Lacs in the 1LAC sample. After excluding the FSRQs, we find essentially the
same result as before. The values for the three bins change by less than $1\sigma$, although because
the FSRQs are predominantly LSP, the uncertainty for the LSP bin increases
substantially due to the reduced number of sources. The HSP population, which is entirely BL~Lacs,
is less variable than either the ISP or LSP populations at the ${4\sigma}$ level. The ISP and LSP
BL~Lac populations differ by less than $1\sigma$.

From our 1LAC sample, we measured intrinsic modulation indices for 258 sources that also have
\nupeak{} values in the 2LAC catalogue~\citep{ackermann_second_2011}. These are plotted in
Fig.~\ref{fig:im_vs_nupeak}. BL~Lacs and FSRQs clearly occupy different regions of this plot, with
FSRQs confined to the left edge but reaching high levels of variability, while BL~Lacs are plentiful
across the full range of \nupeak{} values with only low intrinsic modulation indices at high
\nupeak{} values. We note that the sources with the three highest intrinsic modulation indices,
\mbox{J0238+1636}, \mbox{J0654+4514}, and \mbox{J0050$-$0929}, are all highly compact radio sources
on parsec scales, which likely indicates that these sources are viewed very nearly along their jet
axes~\citep{lister_mojave_2009,lister_gamma-ray_2011,lister_mojave_2013}.

\subsection{Redshift Trend}

\begin{figure}
  \centering
  \includegraphics[width=\columnwidth]{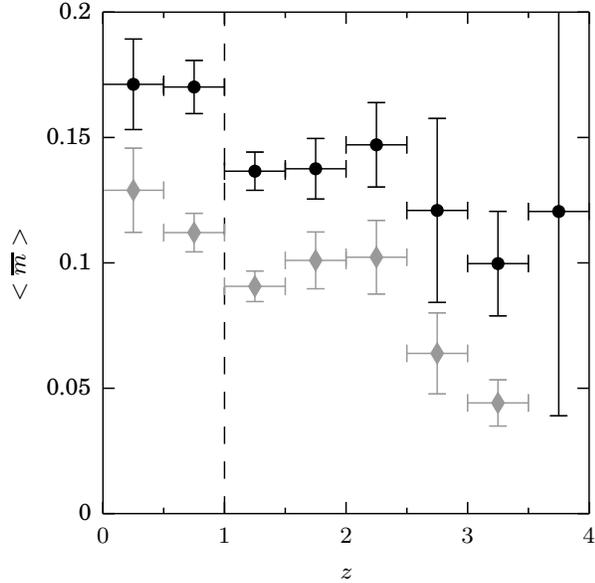}
  \caption{Mean modulation indices for bright ($S_0>400~\mathrm{mJy}$) CGRaBS
    FSRQs in redshift bins with $\Delta z=0.5$. Horizontal error bars indicate bin widths, vertical
    error bars indicate the statistical uncertainty in the plotted means. Black circles indicate
    data points computed from the four~year data set, grey diamonds from the two year data set. The
    vertical dashed line indicates $z=1$.}
  \label{fig:m_vs_z}
\end{figure}

\begin{figure}
  \centering
  \includegraphics[width=\columnwidth]{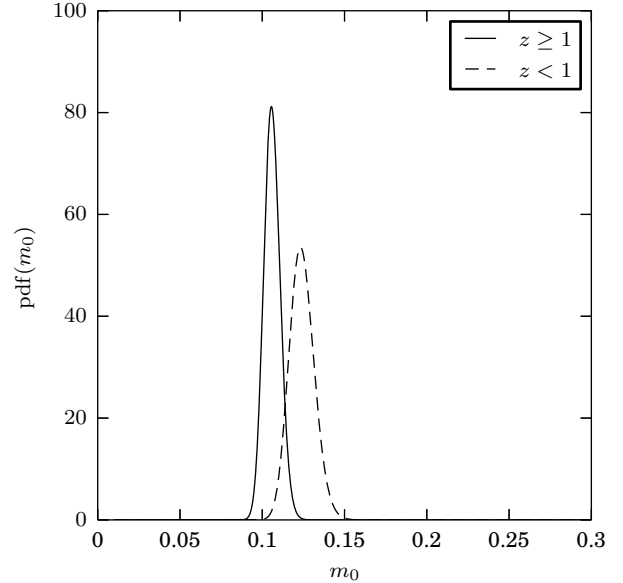}
  \caption{Likelihood distributions for $m_0$ for the CGRaBS FSRQs
    with known redshift in our monitoring sample with $z\geq1$ (solid
    line) and $z<1$ (dashed line). The most likely values for the mean
    modulation index for each distribution and for the difference
    between the two are listed in Tables~\ref{tab:cgrabs_modindices}
    and~\ref{tab:population_results}. The two distributions \emph{are}
    consistent with having the same mean modulation index at the
    $2\sigma$ level.}
  \label{fig:cgrabs_fsrq_hiz_lowz}
\end{figure}

Based on the two-year results, we found evidence that the intrinsic modulation indices for CGRaBS
FSRQs in our sample decreased with increasing redshift~\citep{richards_blazars_2011}. In
Fig.~\ref{fig:m_vs_z}, we plot with black circles the mean four-year intrinsic modulation indices
among bright (${S_0>400~\mathrm{mJy}}$) CGRaBS FSRQs as a function of redshift. Although the trend
suggested by the two-year data, shown here with grey diamonds, remains visible, the scatter of the
data within each bin has increased, particularly at higher redshifts. None the
  less, the correlation is significant (Kendall's ${\tau=-0.15}$ with
${p<5\times10^{-5}}$).  This is not due, e.g., to mutual correlation with flux density. We do find a
significant correlation between redshift and $S_0$ (${\tau=-0.15}$, ${p<5\times10^{-3}}$) but no
correlation between $S_0$ and intrinsic modulation index (${\tau=0.04}$, ${p<0.48}$). In
Fig.~\ref{fig:cgrabs_fsrq_hiz_lowz} we plot the likelihood distributions for CGRaBS FSRQs at high
($z\geq1$) and low ($z<1$) redshift, including those fainter than 400~mJy. Although we continue to
find a larger most-likely value of $m_0$ for FSRQs at lower redshift (most likely difference
\valunc{0.018}{0.009}), the significance of the separation is less than $2\sigma$ and has fallen
slightly compared to the two-year result.

As discussed in \citet{richards_blazars_2011}, there are several competing effects that will
contribute to such a trend. Due to cosmological time dilation, equal-length observations will
correspond to different source-frame time intervals according to ${\Delta
  t_{\mathrm{source}}=\left(1+z\right)^{-1}\Delta t_{\mathrm{obs}}}$. Because, as we showed in
section~\ref{sec:variability_trend}, the intrinsic modulation index increases with the observation
length, this will tend to decrease $\im$ with increasing redshift. This effect explains at least
part of our observed redshift trend. \citet{jlr_thesis} used a subset of the OVRO data set to
investigate this effect by comparing the variability in equal rest-frame time periods. Accounting
for this reduced the most likely difference between the ${z\geq1}$ and ${z<1}$ samples slightly,
although the change was within the uncertainty.

At higher redshifts, the 15~GHz observed frequency corresponds to a higher rest frame emission
frequency. Although blazar variability indices do not differ significantly between 8~GHz and 90~GHz,
the source-frame characteristic time between blazar flares at 15~GHz is somewhat
longer than at 37~GHz~\citep{hovatta_statistical_2007, hovatta_long-term_2008}. Thus, lower
redshift blazars will likely have undergone fewer flares in a given source-frame time period. This
will tend to reduce the observed intrinsic modulation index, particularly before the source's range
of behavior has been completely measured. The overall increase in intrinsic modulation
indices between the two- and four-year data sets, illustrated in
Fig.~\ref{fig:modindex_scatter}, suggests that we have not yet achieved this. This will tend to
increase the intrinsic modulation index with redshift, in opposition to our detected trend.

Other effects, such as Doppler and Malmquist biases will also contribute to the redshift
trend~\citep{lister_statistical_1997}. Determining whether an intrinsic component to this apparent
evolution is present will require Monte Carlo simulation to assess the net contributions of the
various biases.

\section{Discussion}

We have presented and examined the results of four years of twice-weekly 15~GHz radio monitoring of
more than 1400 blazars, including systematically radio- and gamma-ray-selected samples. Using the
intrinsic modulation index to quantify the variability amplitude for each source, we find that on
average, gamma-ray-loud CGRaBS sources are significantly more radio variable than gamma-ray-quiet
CGRaBS sources, confirming our previous result~\citep{richards_blazars_2011}. This reflects a
significant difference in variability between FSRQs in the radio-selected CGRaBS and
gamma-ray-selected 1LAC samples. This indicates that among FSRQs, there is clearly a connection
between gamma-ray emission and radio variability. The 1LAC sample contains primarily strongly
radio-variable FSRQs. The CGRaBS sample also contains many FSRQs that are both gamma-ray-loud and
strongly radio-variable, but it also contains a substantial gamma-ray-quiet, weakly radio-variable
FSRQ population. BL~Lac objects show similar radio variability in both the gamma-ray- and
radio-selected samples.

Within the 1LAC sample, the radio variability amplitude is connected to the frequency of the
synchrotron SED peak, $\nupeak$. HSP blazars, with ${\nupeak>10^{15}~\mathrm{Hz}}$, show
significantly less radio variability than do either ISP or LSP blazars. This result persists when
FSRQs are excluded and only BL~Lac objects are considered. In fact, the variability of both the ISP
and LSP BL~Lacs is the same as that of FSRQs to the $1\sigma$ level. This agrees with and quantifies
a previous report that HSP BL~Lac objects tend to have moderately low modulation indices in the OVRO
15~GHz data, while many LSP and ISP sources vary at higher levels~\citep{lister_gamma-ray_2011}. We
do not find evidence for a difference in radio variability amplitude between the ISP and LSP
blazars, and among BL~Lacs find a slightly lower level of variability among LSPs, although the
uncertainties are relatively large and the difference is less than $1\sigma$.

\citet{hovatta_connection_2013} examined blazar optical variability and found a monotonic increase
from HSP to ISP to LSP sources. A continuous trend in gamma-ray variability index versus $\nupeak$
was also found by \citet{ackermann_second_2011}. The ability to detect such a trend in these bands
benefits from the substantially stronger variability there than at 15~GHz. In gamma-rays in
particular, sources are typically undetected when quiescent, whereas the presence of extended radio
emission provides a steady background that dilutes the fractional variability measured by the
intrinsic modulation index. We find that the steady flux density background increases at lower
$\nupeak$ values. If there is an upper limit to the component of the flux densities that is variable
on the week-to-year time-scales we probe with this programme, this would lead to a saturation effect:
as the steady flux density increases, the intrinsic modulation index corresponding to the maximum
variable flux density will decrease.  A limit on the variable flux density is likely because we
expect limits on intrinsic brightness temperatures~\citep{readhead_equipartition_1994,
  kellermann_spectra_1969}. 

These results suggest that, among FSRQs, there is a connection between detectable gamma-ray emission
and radio variability. No evidence for such a connection is found for BL~Lacs. This finding is
similar to the results of several other recent radio blazar studies.  Compared to their
gamma-ray-quiet counterparts, LAT-detected gamma-ray-loud FSRQs have been found to exhibit faster
parsec-scale jet speeds~\citep{lister_connection_2009} and larger core brightness temperatures,
incidence of polarization, and parsec-scale opening angles~\citep{linford_characteristics_2011,
  linford_contemporaneous_2012}. In most cases, no significant difference has been found between
gamma-ray-loud and gamma-ray-quiet BL~Lac populations. \citet{linford_characteristics_2011} did
report gamma-ray-loud BL~Lacs to have longer jet lengths and \citet{linford_contemporaneous_2012}
found they had a slightly higher incidence of core polarization.

The presence of this connection in FSRQs and its apparent absence in BL~Lacs may result from
different sources of inverse Compton seed photons in these classes. High-energy photons produced by
the EC process are beamed more strongly compared to photons produced by either the synchrotron or
SSC processes~\citep{dermer_beaming_1995}, so for a single Lorentz-factor jet, gamma rays produced
via EC will be emitted into a smaller solid angle than SSC gamma rays.  Several studies have found
evidence that BL~Lac sources are typically dominated by the SSC process while FSRQs chiefly emit via
the EC process~\citep[e.g.,][]{abdo_spectral_2010, lister_gamma-ray_2011}, and it seems that more
powerful jets like those found in FSRQs are more likely to exhibit
EC~\citep{meyer_collective_2012}. There do, however, seem to be exceptions to this
rule~\citep[e.g.,][]{boettcher_modeling_2012}. Still, if this connection is generally correct, then
the gamma-ray-loud subset of FSRQs would consist of those whose line of sight falls within the
narrow gamma-ray emission cone, which would be the higher-Doppler factor sources. Higher Doppler
factors are associated with increased radio
variability~\citep[e.g.,][]{lahteenmaki_and_valtaoja_1999, jorstad_superluminal_motion_2001,
  hovatta_doppler_2009}, so this would lead to gamma-ray-loud FSRQs being more radio variable than
the overall average, as we find. In BL~Lacs, on the other hand, SSC gamma-ray
emission would be subject to the same beaming factor as the radio emission, so this selection bias
toward higher Doppler factors will be weaker or absent. The population of BL~Lacs that exhibit
substantial gamma-ray emission would primarily be determined by factors other than beaming effects,
as suggested by~\citet{lister_connection_2009}.

We caution, however, that we may be over-interpreting our BL~Lac result: a selection effect may bias
our variability average for this class. The CGRaBS and 1LAC BL~Lac samples differ strongly in the
proportion of HSP sources in each. Although we do not have uniform SED classifications for the
CGRaBS sample, of the 60 that do have classifications only 11 are HSP. For 1LAC, 85 of the 168 with
classifications are HSP. Since HSP BL~Lacs are much less variable than either ISP or LSP BL~Lacs, it
is surprising that we find overall similar intrinsic modulation indices for the CGRaBS and 1LAC
BL~Lac samples. This happens because HSP sources are fainter and less variable in radio, and so are
more likely to be dropped by our cuts in $S_0$ and $\im$. As a result, HSP BL~Lacs are
under-represented in the determination of the mean intrinsic modulation index for the 1LAC sample,
leading to an overestimate of $m_0$.

Finally, the observed trend of decreasing variability amplitude with increasing redshift among
CGRaBS FSRQs appears still to be present in our longer data sets, albeit with reduced significance
when we compare the mean intrinsic modulation indices of high and low redshift objects. The
correlation between $\im$ and $z$ for bright FSRQs is highly significant (${p<5\times10^{-5}}$),
however. There are several effects and biases that will contribute to such a trend, so determining
whether a cosmological evolution component is present will likely require simulations.

\section*{Acknowledgements} JLR would like to thank Matthew~L. Lister for several helpful
  discussions. The OVRO 40~m monitoring programme is supported in part by NSF grants AST-0808050 and
AST-1109911 and NASA grants NNX08AW31G and NNX11AO43G. TH was supported by the Jenny and Antti
Wihuri foundation. VP acknowledges support from the `RoboPol' project, which is implemented under
the `Aristeia' Action of the `Operational Programme Education and Lifelong Learning' and is
co-funded by the European Social Fund (ESF) and Greek National Resources; and from the European
Commission Seventh Framework Programme (FP7) through grants PCIG10-GA-2011-304001 `JetPop' and
PIRSES-GA-2012-31578 `EuroCal.' The National Radio Astronomy Observatory is a facility of the
National Science Foundation operated under cooperative agreement by Associated Universities, Inc.
This research has made use of the NASA/IPAC Extragalactic Database (NED) which is operated by
  the Jet Propulsion Laboratory, California Institute of Technology, under contract with the
  National Aeronautics and Space Administration.

\footnotesize{
  \bibliography{ms}
}

\end{document}